\documentclass{article}

\usepackage[utf8]{inputenc}
\usepackage{color}
\usepackage{graphicx}
\usepackage{amsmath}
\usepackage{booktabs}
\usepackage{mathtools}

\usepackage[authoryear]{natbib}

\usepackage{xr}

\makeatletter
\newcommand*{\addFileDependency}[1]{
  \typeout{(#1)}
  \@addtofilelist{#1}
  \IfFileExists{#1}{}{\typeout{No file #1.}}
}
\makeatother

\newcommand*{\myexternaldocument}[1]{%
    \externaldocument{#1}%
    \addFileDependency{#1.tex}%
    \addFileDependency{#1.aux}%
}

\makeatletter
\long\def\XR@test#1#2#3#4\XR@{%
  \ifx#1\newlabel
    \mod@newlabel{\XR@prefix#2}#3{}{}\@nil
  \else\ifx#1\@input
     \edef\XR@list{\XR@list#2\relax}%
  \fi\fi
  \ifeof\@inputcheck\expandafter\XR@aux
  \else\expandafter\XR@read\fi}
\def\mod@newlabel#1#2#3#4\@nil{\newlabel{#1}{{#2}{#3}}}
\makeatother

\DeclarePairedDelimiter\floor{\lfloor}{\rfloor}

\providecommand{\covid}{SARS-CoV-2}
\providecommand{\covids}{SARS-CoV-2 }
\providecommand{\rtm}{RTM}
\providecommand{\rtms}{RTM }

\providecommand{\logit}{\textrm{logit}}
\renewcommand{\vec}[1]{\boldsymbol{{#1}}}
\providecommand{\myth}[1]{${#1}^{\textrm{th}}$}
\providecommand{\myst}[1]{${#1}^{\textrm{st}}$}

\providecommand{\nni}{\Delta}

\providecommand{\ksens}{k_{\textrm{sens}}}
\providecommand{\kspec}{k_{\textrm{spec}}}
\newcommand{\R}{\mathcal{R}}

\newcommand{\Sstate}[2]{S^{V_{{#1}}}_{{#2},a}}

\newcommand{\Istate}[3]{I^{V_{{#2}},{#1}}_{{#3},a}}
\newcommand{\Istateprime}[3]{I^{V_{{#2}},{#1}}_{{#3},a'}}
\newcommand{\Rstate}[3]{R^{V_{{#2}},{#1}}_{{#3},a}}
\newcommand{\Wstate}[3]{W^{V_{{#2}},{#1}}_{{#3},a}}

\myexternaldocument{Appendix}

\title{Real-time modelling of the SARS-CoV-2
pandemic in England 2020-2023: a challenging
data integration}
\author{Paul J Birrell$^{1,2*}$, Joshua Blake$^2$, Joel Kandiah$^2$, Angelos \\ Alexopoulos$^{2,3}$, Edwin van Leeuwen$^1$, Koen Pouwels$^{4,5}$, Sanmitra\\ Ghosh$^2$, Colin Starr$^2$, Ann Sarah Walker$^{6,5}$, Thomas A House$^7$,\\ Nigel Gay$^8$, Thomas Finnie$^1$, Nick Gent$^9$, Andr\'e Charlett$^1$,\\ Daniela De Angelis$^{2,1}$}
\date{}

\begin{document}

\maketitle

\begin{itemize}
\item[$^1$] Data, Analytics and Surveillance, UK Health Security Agency, 61 Colindale Avenue, London NW9 5EQ, UK
\item[$^2$] MRC Biostatistics Unit, University of Cambridge, East Forvie Building, Forvie Site, Robinson Way, Cambridge CB2 0SR, UK
\item[$^3$] Department of Economics, Athens University of Economics and Business, Greece.
\item[$^4$] Health Economics Research Centre, Nuffied Department of Population Health, University of Oxford, Oxford OX3 7LF, UK
\item[$^5$] The National Institute for Health Research Health Protection Research Unit in Healthcare Associated Infections and Antimicrobial Resistance, University of Oxford, Oxford OX3 7LF, UK
\item[$^6$] Nuffield Department of Population Health, University of Oxford, Oxford OX3 7LF, UK
\item[$^7$] Department of Mathematics, University of Manchester, Oxford Road, Manchester M13 9PL, UK
\item[$^8$] Fu Consulting, Hungerford, UK
\item[$^9$] Chief Medical Officer, Cayman Island Government, 133 Elgin Avenue, Grand Cayman KY1-9000, Cayman Islands
  \medskip\medskip
  \item[$^*$] \textit{Corresponding author:} paul.birrell@ukhsa.gove.uk
\end{itemize}

\section*{Abstract}

A central pillar of the UK’s response to the SARS-CoV-2 pandemic was the provision of up-to-the moment nowcasts and  short term projections to monitor current trends in transmission and associated healthcare burden. Here we present a detailed deconstruction of one of the `real-time’ models that was key contributor to this response, focussing on the model adaptations required over three pandemic years characterised by the imposition of lockdowns, mass vaccination campaigns and the emergence of new pandemic strains. The Bayesian model integrates an array of surveillance and other data sources including a novel approach to incorporating prevalence estimates from an unprecedented large-scale household survey. We present a full range of estimates of the epidemic history and the changing severity of the infection, quantify the impact of the vaccination programme and deconstruct contributing factors to the reproduction number. We further investigate the sensitivity of model-derived insights to the availability and timeliness of prevalence data, identifying its importance to the production of robust estimates.

\textbf{Keywords}: transmission modelling, reproduction number, severity estimation, prevalence survey, Bayesian melding, nowcasting

\section{Introduction}

The WHO declared the outbreak of \covid\ to be a global pandemic on \myth{11} March, 2020. Prior to this date, and over the ensuing three years, many countries introduced  mitigation measures designed to limit the healthcare burden and loss of life due to the pandemic. The most severe measures included societal `lock-downs', limiting access of many citizens to their places of work and education, friends and family, and many basic services, while new vaccines and vaccine technologies were developed and distributed at an unparalleled rate. In the United Kingdom (UK), there were three national-level lock-downs: from the \myth{23} March, 2020 to \myth{4} July, the most restrictive lockdown; from \myth{5} November to \myth{2} December 2020; and a third lockdown beginning on the \myth{6} January 2021. Relaxation of this final lockdown was a staged and gradual process with most restrictions having been removed by \myth{19} July 2021.

The decisions to impose and relax lock-downs, alongside the introduction of many other interventions 
were extensively underpinned by outputs from various mathematical and statistical models. In the UK, the most high-profile modelling informing government policy was carried out under the auspices of the Scientific Pandemic Influenza Subgroup on Modelling (SPI-M), a sub-group of the Scientific Advisory Group on Emergencies (SAGE) comprising academic teams and staff drawn from Public Health England (PHE, later the UK Health Security Agency, UKHSA) and the UK Department of Health and Social Care (DHSC). SPI-M monitored the pandemic's ongoing threat through two tasks: nowcasting and forecasting. Nowcasting involved the estimation of a number of key indicators, primary among which were reproduction numbers \citep{Pellis2022}, for which a value below unity is indicative of declining infection transmission; and daily numbers of both incident and prevalent infections. Estimates of these indicators were available both nationally and for each of the seven National Health Service (NHS) regions of England. Forecasting involved the production of medium-term projections (MTPs) of measures of severe disease burden over the coming eight week period under an assumption of no change in policy or public behaviour. They included the number of new positive cases identified in hospitals, total hospital bed occupancy by test-positive individuals, and deaths. For both nowcasts and MTPs, outputs from contributing modelling efforts were combined via model stacking methods to give consensus estimates and projections that encompassed the uncertainty both within and between modelling team's outputs \citep{Maishman2022, Silk2022, Park2023}.
The cohort of models used included mechanistic approaches, where transmission of the virus is explicitly modelled through a SEIR-type ({\bf S}usceptible--{\bf E}xposed--{\bf I}nfected--{\bf R}ecovered) model, which partition populations into disease states \citep{c18ef097-75bd-3f59-a5d7-256a8f9b744e}; and other, semi-mechanistic, approaches relying on renewal or time-since-infection-type models that relate incident infections to the recent history of infection \citep{Cori2013}.  A selection of the mechanistic approaches can be found in \citep{Overton2022,Perez-Guzman23,Keeling2021,Keeling2022} with the semi-mechanistic approaches exposed in \citep{Mishra2022, Scott2021, Abbott2020, Abbott2022, WhyTeh22, Ackland2022}.

In particular, the PHE-Cambridge Real-Time Model (\rtm) contributed outputs to policymakers from mid-March 2020. Developed pre-pandemic \citep{BirKGCPetal09,BirPCZD17} as a deterministic, compartmentalised model, it was rapidly deployed following the outbreak of \covid\ to capture the dynamics of the new outbreak. Over the following three years, continual adaptation and extension were required to deal with a long-lived pandemic subject to unprecedented levels of public-health intervention and surveillance. In \citet{BirBVGD21}, we presented an earlier, simpler version, of the model, sufficient to tackle the initial wave of infection and the first lockdown. Here we give full details of how the initial model had to be progressively developed to incorporate novel data sources to address the acute challenges posed by changes in the mix of circulating variants; vaccination; reinfection (after the emergence of the Omicron variants); as well as a periodic sparsity of surveillance information. 
 
Extending previous work \citep{BirBVGD21}, here we detail a Bayesian approach to inference that permitted the timely estimation of latent features of the pandemic 
 over the full three years from March 2020 to March 2023.  
 To do so it was necessary to assimilate data from multiple sources of different size and quality, which
 required addressing new complexities that the synthesis \citep{DeaPBSH15,Birrell2018} of heterogeneous data, only indirectly informing quantities of interest and accumulating over such a long time brings in terms of both modelling and inference. These complexities include: specification of appropriate observational models for data held on secure and remote trusted research environments (TREs), quantifying changes to the transmissibility of a virus through a stochastic process, and the consequent 
development of bespoke computational algorithms to sample from a posterior distribution of progressively increasing dimension. 

The paper is organised as follows: in Section \ref{sec:data} the data sources used in the \rtms are reviewed and the model developments are introduced chronologically; in Section \ref{sec:results} we present routine outputs 
that the model provided over the course of the pandemic, together with additional insights, which can also be produced in real-time, 
such as decomposition of the reproduction number to understand drivers of transmission and estimates of the impact of the vaccination programme; and in Section \ref{sec:disc2}, we consider the sensitivity of model outputs to the inclusion of a specific data stream, illustrating the fundamental need for an evidence synthesis of this nature to evaluate the role and coherence of different data sources. We further discuss some of the compromises that had to be made due the time pressures exerted by the need of rapid pandemic response, and, relatedly, conclude by suggesting future work, essential to ensuring future pandemic preparedness.

\section{Data and Methods}\label{sec:data}

  \subsection{Data}\label{sec:prev.data}
    In the study of the first wave of infection up to \myth{19} June, 2020 \citep{BirBVGD21}, inference was derived on the basis of: UKHSA's line-listing of all-cause deaths reported to have occurred within 60 days of a lab-confirmed positive reverse transciptase-polymerase chain reaction (RT-PCR) test for the presence of \covids infection; serological data on the proportion of blood samples submitted to the NHS Blood and Transplant (NHSBT) service that tested positive for the presence of various antibodies (detailed below); and mobility indices derived from Google mobility, the UK Time-Use survey (UKTUS) and data on school attendances from the UK Department for Education (DfE) \citep{vLeS22}. Here the same data sources are retained, stratified by seven NHS regions and eight age groups ($<1$, 1--4, 5--14, 15--24, 25--44, 45--64, 65--74, $\geq75$). 
     Additionally, we used information on the timing and number of administered vaccine doses, essential to understand how exposed the population was to successive waves of infection and severe disease.
     
       However, deaths became very sparse during the summer of 2020, with fewer than five deaths nationwide on \myth{19} August, identifying a need to augment data on deaths with additional and alternative sources of information to inform the inference. 
Data on the daily numbers of newly diagnosed cases could have also been used. However, trends in these data were subject to biases that resulted from changes to test availability and testing behaviours in response to rapidly changing governmental policy, including `surge' testing, a common practice at infection hot-spots or during the emergence of new variants \citep{NicLPPJ21}. Therefore, we used information from the Office for National Statistics (ONS) COVID-19 Infection Survey (CIS) and data on hospital admissions, as an alternative to deaths data (which could also be influenced by testing propensity). Detailed description of the active data sources are given in what follows.

\paragraph{Prevalence} The CIS recruited randomly-selected private households on a continual basis from \myth{26} April 2020 to \myst{31} January 2022 to provide a representative sample of households across the UK. Participating household members two years of age and over were routinely tested for \covids infection using RT-PCR tests, initially at weekly intervals over a period of four weeks, then at monthly intervals over a period of a year. Follow up of participating households continued until March 2023 \citep{Wei2023}. Use of CIS data is not without challenges: (i) there was a requirement for data generated by CIS to be stored and analysed within the ONS's SDE, the Secure Research System, which had limited computational capacity; (ii) while the invited population was truly a random sample, certain subgroups of the population may be less likely to participate, despite financial compensation for participation. Informative disclosure is avoided by limiting data extraction from the ONS TRE: it would not be possible to extract raw data on the daily number of tests and positive tests at the required level of stratification. Instead, we routinely generated daily estimates of the age and NHS-region specific number of \covids infections that would test RT-PCR-positive, a measure of infection prevalence. These estimates were derived from a Bayesian multilevel regression and poststratification (MRP) procedure designed to ameliorate any effects of a lack of representativeness in the study. The MRP approach is an adaptation of the method of \citet{PouHPetal21}, designed to correct estimates of positivity for any potential sampling biases in terms of location, age, sex and time (see Section C 
of the Online Appendix for a detailed description of how these estimates are generated). 

\paragraph{Serology} While serological data, as detailed in \citet{BirBVGD21}, were used for the first wave, there was evidence of waning antibody positivity detected by the EuroImmun SARS-CoV-2 ELISA IgG assay. From \myth{25} November 2020 onwards, NHSBT also tested samples using a nucleocapsid assay from Roche (the `Roche-N' assay), which has an enduring antibody response capable of detecting prior infection, but not prior vaccination \citep{Whitaker2022}, thus providing a measure of the size of the uninfected population. In total we include in our analysis 13,478 samples between \myth{26} March and \myst{21} May 2020 tested using the EuroImmun assay, and a further 216,243 samples between \myth{25} November 2020 and \myth{17} March 2023 tested using the Roche-N assay. All samples were then assigned a date 25 days later than the date the sample was taken to account for the time required for an antibody response to develop.

    \paragraph{Vaccinations}
    Daily data on the numbers of people being immunised, stratified by age-group, region, dose number and vaccine type were collected by the National Immunisation Management Service (NIMS). These data include all \covid\ immunisations administered at hospital hubs, local immunisation service sites such as GP practices, and dedicated immunisation centres, but does not count any vaccinations obtained abroad. The vaccination campaign began on December \myth{8}, 2020 with a 2-dose primary vaccination schedule. Third and fourth `booster' doses were made subsequently available from \myth{16} September 2021 and \myth{7} September 2022 respectively. Like the serological samples, the vaccination data were assigned a date 21 days later than the date of the immunisation to allow for an immune response to develop \citep{Hall2021vacc}.

    \paragraph{Admissions}
    To provide the insight needed to understand the demands \covid\  was placing across the health sector and to coordinate the appropriate allocation of resources and services, the NHS, in March 2020, launched a collection of Situation Reports (SitReps). The COVID-19 Daily NHS Provider SitRep required NHS hospital trusts to report, amongst other things, the number of new diagnoses among admitted patients. These new diagnoses are stratified into age classes and by the time between admission and the positive test \citep{NHSDigital22}. We constructed a time series of hospital admissions due to \covids infection by counting only those positive tests recorded by the hospital trusts as having been detected within two days of admission. The SitRep data do not, however, extend into the pre-lockdown era, with the first report published on the \myth{19} March, 2020. Therefore, the SitRep admissions are augmented by the NHS England Secondary Uses Service (SUS) dataset, which contains complete and accurate information on hospitalisations for \covid\  in England. Admissions are, however, only entered into the SUS dataset upon completion of a hospital stay (i.e. at the point of discharge from hospital or death) and as such are a heavily lagged dataset. Though the historical SUS and SitRep data are qualitatively similar, they do not correspond exactly. We therefore use a hybrid hospital admission dataset comprised of the SUS data up to \myth{5} May 2021 and the SitRep data thereafter. The date at which the datasets are knitted together was chosen to be the date at which the two datasets most closely correspond over a sustained period.

    \subsection{Methods}\label{sec:methods}

\subsubsection{Modelling Background}

The starting point for this work is the model developed to reconstruct the transmission dynamics of the \covids  pandemic in England introduced in \citet{BirBVGD21}. The model dynamics are laid out in Appendix Section A, 
with mathematical and graphical descriptions of the model found in Equation (A.1) 
and Figure A.1 
respectively. In summary, $S_{r, t_k, a}$, $E^l_{r, t_k, a}$, $I^l_{r, t_k, a}, l = 1, 2$ represent the number of people in the $S$ (susceptible), $E$ (exposed, not infectious), $I$ (infectious) and $R$ (recovered/removed) disease states that partition the population at time $t_k$, in region $r, r = 1, \ldots, n_r$ and age group $a, a = 1, \ldots, n_a$. The model is evaluated at discrete timepoints, $t_k = k\delta t$, $k = 1, \ldots, K$, such that the \myth{k} time interval is $((k-1)\delta t, k\delta t]$ with time-step $\delta t$ chosen to be 0.5 days.

New infections $\nni_{r,t_k,a}$ are generated through the interaction of susceptible and infectious individuals as
  \begin{equation}\label{eq:nni}
    \nni_{r,t_k,a} = S_{r,t_k,a}\lambda_{r,t_k,a}\delta t,
  \end{equation} where
\begin{equation}\label{eqn:prob.infection}
    \lambda_{r,t_k,a} = \left( 1 - \prod_{a'=1}^{n_a} \left[ \left( 1 - b^{t_k}_{r,aa'} \right)^{I^1_{r,t_k,a'} + I^2_{r,t_k,a'}} \right] \right).
\end{equation}
is the time- and age-varying the force of infection, {\it i.e.} the rate with which susceptible individuals become infected, expressed in terms of $b^{t_k}_{r,aa'}$,  the probability of a susceptible individual in region $r$ of age group $a$ being infected by an infectious individual in age group $a'$ at time $t_k$. This probability is a function of: a set of time-varying contact matrices, $\vec{C}^{t_k}$, describing the rates of contact between individuals of different age groups over time (see \citet{vLeS22} and Section A 
of the Online Appendix); parameters, $m_{r,a}$, modifying the contact matrices that have the interpretation of age-specific (relative) susceptibilities to infection given contact with an infectious individual; time-varying transmission intensities, $\beta_{r,t_k}$, quantifying the temporal changes in the virus transmissibility, incorporating all un-modelled factors (virological, environmental, behavioural etc); the initial growth rate in the number of infections in each region, $\psi_r$; and the mean duration of infectiousness $d_I$ (see Section A 
of the Online Appendix for details). 

In what follows we detail how this model has been adapted to rise to the challenges of an ever-shifting pandemic landscape.

\subsubsection{Model Adaptations}

\paragraph{August--December 2020 --- addressing data sparsity}

The decrease in number of deaths over the summer 2020 indicated the need for alternative sources of data and highlighted changes in the risk of death. Addressing both of these led to model developments.

\begin{figure}
  \centering
    (A)\\
    \includegraphics[width=0.8\linewidth]{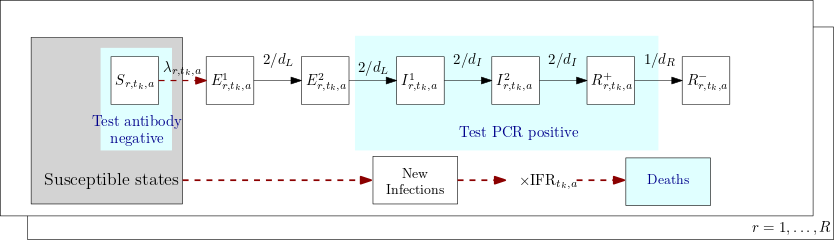}\\
    (B)\\
    \includegraphics[width=0.8\linewidth]{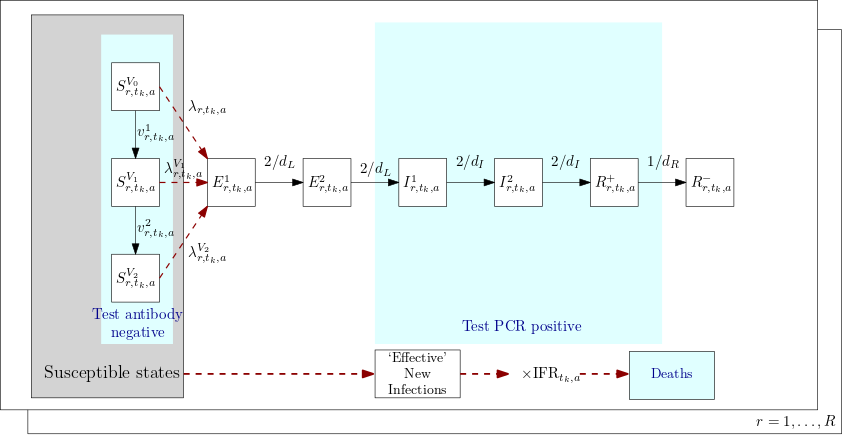}\\
    (C)\\
    \includegraphics[width=0.8\linewidth]{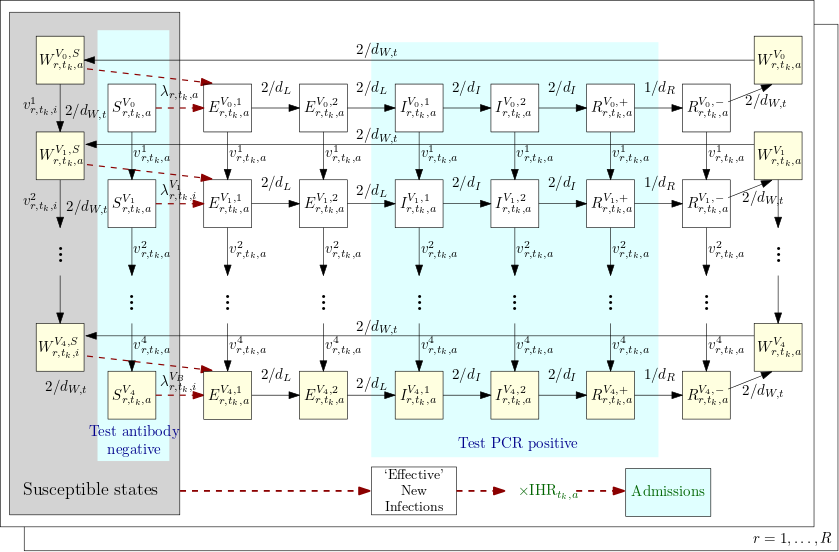}
    \caption{Model adaptations throughout the pandemic: (A) addition of ONS CIS information; (B) stratification of susceptible states by vaccination dose; (C) stratification of the entire model by vaccine dose to account for waning and expansion to account for the booster vaccination programme. Shaded blue areas are `observed' quantities, grey-shaded areas enclose susceptible individuals/groups.}
    \label{fig:mod.figs}
\end{figure}

Incorporation of the ONS CIS prevalence estimates as an additional data stream required a change in model structure to to track the prevalence of RT-PCR positive infection. This was achieved by partitioning the (R) recovered state in the original SEEIIR model into two states R$^+$ and R$^-$ (see Figure \ref{fig:mod.figs} (A)), with individuals in R$^+$ still testing positive (subject to the sensitivity of the test), but no longer being infectious. The average time spent in this state is denoted $d_R$. This partition accounts for the expected duration of RT-PCR positivity being longer than the infectious period \citep{SinPCBSELZG20}.  

A further benefit of adding CIS data has been the ability to inform the regional susceptibility-to-infection parameters introduced  
in \citet{BirBVGD21}, to describe differential susceptibility for the over-75s, the effect of the lockdown and the interaction term between these two factors. The CIS information permitted the estimation of these susceptibility parameters with increased age specificity (data on deaths are disproportionately informative on older age groups), moving from three to eleven per region. The contribution of these parameters is summarised around Equation 
(A.5) in the Web Appendix.

To address the changing risk of dying from \covids \citep[e.g.][]{Kirwan2022}, a time-varying infection fatality ratio (IFR, the fraction of infections that will result in a \covid-associated death), a quantity previously assumed to be constant over time (in \citet{BirBVGD21}), was introduced in the model. More specifically:
the number of severe events, $\mu_{r,t_k,a}$, are derived from the scaled convolution,
\begin{equation}\label{eqn:convolutionA}
  \mu_{r,t_k,a} = \sum_{l=0}^k f_{k - l} p_{t_l, a} \nni_{r, t_l, a},
\end{equation}
where $f_k$ are quantiles of the probability distribution governing the time from infection to the severe event, and $p_{t_l, a}$ are age-specific severity ratios. When dealing with data on deaths, this ratio is the IFR. Denote $\vec{\zeta}_a$ to be age-specific parameters that adjust the levels of the IFR at changepoints introduced approximately every 100 days.
The transition between levels of the IFR are assumed to take place linearly on the logistic scale over the course of thirty days, to avoid sudden jumps in the expected number of observed deaths. Formally, if there were $s$ changepoints at times $t'_1, \ldots, t'_s$, then
\begin{equation*}
\logit\left(p_{t_k, a}\right) = \sum_{s' = 1}^s g\left(\frac{t_k - t'_{s'}}{30}\right)\zeta_{a,s'}
\end{equation*}
where
\begin{equation*}
    g(x) = \begin{cases}
    0 &  x<0\\
    x & x \in \left[0,1\right]\\
    1 & x > 0
    \end{cases}.
\end{equation*}

\paragraph{December 2020 -- May 2021 --- launch of the vaccination campaign} 

The immunisation campaign served to reduce the healthcare burden associated with the pandemic, directly, by preventing infected individuals from developing severe illness and, indirectly, by interrupting transmission and reducing the spread of infection within the population. Vaccination is often assumed to be either `all-or-nothing', in which case a fraction of immunised individuals have fixed and lasting protection against infection or disease, or it is `leaky', i.e. vaccinated individuals can still become infected, albeit with reduced likelihood \citep{McLean1995,Arino2004,Wwick_vaccine_waning}. Here, we make a number of assumptions: that the vaccine provides leaky protection against infection 
\citep{UKHSA21,Zachreson2023} that once infected, disease transmission and infection duration are independent of vaccination status; and that the probability of being vaccinated is independent of disease state. Figure \ref{fig:mod.figs}(B) illustrates how the model was adapted to stratify the susceptible population by vaccine dose to account for the differential infection and severity risk in vaccinated individuals. In the figure, the super-script $V_q$ denotes a state or quantity referring to individuals who have received $q$ vaccine doses.
The $v^q_{r,t_k,a}$ transition rates are the observed rates of vaccination with a \myth{q} dose. If, on day $d$, there are $V^q_{r,d,a}$ people in region $r$ and age-group $a$ who have newly received a \myth{q} vaccine dose, out of a population of size $N_{r,a}$, then to translate these into the rates required by the model we need to know what fraction of the susceptible population are being vaccinated. The denominator population for receiving the vaccines in region $r$, age group $a$, dose $q$, and day $d$ are of size:
    \begin{equation*}
        N^q_{r,d,a} = \begin{cases} N_{r,a} - \sum^{d-1}_{l = d_0} V^1_{r,l,a} & q=1\\
          \sum^{d-1}_{l = d_0} V^{q-1}_{r,l,a} - \sum^{d-1}_{l=d_0} V^q_{r,l,a} & q > 1
          \end{cases}
    \end{equation*}
    where $d_0$ is the day on which the vaccination programme was initiated.
    The observed daily fractions of newly vaccinated individuals are
    \begin{equation}\label{eq:vac.obs}
        v^{*q}_{r,d,a} = \frac{V^q_{r,d,a}}{N^q_{r,d,a}}
    \end{equation}
    The model has $\delta t$ time-steps, not daily time steps, with $1/\delta t$ being integer. If we assume that $v^q_{r,t_k,a}$ is constant over all time steps on the same day ($v^q_{r,t_k,a} \equiv v^q_{r,d(t_k),a}$), the model calculates a fraction vaccinated in a day to be (one minus the probability of not being vaccinated on the day):
    \begin{equation}\label{eq:vac.mod}
        v^{*q}_{r,t_k,a} = 1 - \left(1 - v^q_{r,d,a}\delta t\right)^{1/\delta t}
    \end{equation}
    Rearranging for $\delta t = 0.5$ days:
    \begin{equation*}
        v^q_{r,t_k,a} = 2\left(1 - \sqrt{1 - v^{*q}_{r,t_k,a}}\right)
    \end{equation*}
    
The efficacy of the immunisation programme was measured by two time-varying quantities: $\pi^q_{r,t_k,a}$, describing the efficacy of $q$-doses of vaccine at time $t_k$ on age-group $a$ at preventing infection; and $\alpha^q_{r,t_k,a}$, describing the efficacy of $q$-doses of vaccine at preventing the onset of severe illness (either death or hospitalisation, depending on the data in use). Information was recorded on vaccine type, which we classify into two categories, mRNA (either the Pfizer BioNTech BNT162b2 or Moderna vaccines) and non-mRNA (the ChAdOx1-S Astra Zeneca vaccine). The model is not at the individual level, so we cannot track which vaccines each individual has received, but instead we use a weighted average of the assumed mRNA and non-mRNA vaccine efficacies with the weights being equal to the cumulative number of vaccinations given up to that time to a particular region and age-group. 
  For example, for the efficacy against infection
    \begin{equation}\label{eq:efficacy}
        \pi^q_{r,t_k,a} = w^q_{r,t_k,a}\pi^{q,\textrm{mRNA}}_{t_k} + \left(1 - w^q_{r,t_k,a}\right)\pi^{q,\textrm{AZ}}_{t_k}
    \end{equation}
    where
    \begin{equation*}
        w^q_{r,t_k,a} = \frac{\sum_{d=1}^{\floor*{t_k/\delta t}} V^{q,\textrm{mRNA}}_{r,d,a}}{\sum_{d=1}^{\floor*{t_k/\delta t}} V^q_{r,d,a}}
    \end{equation*}
    and  $V^{q,\textrm{mRNA}}_{r,d,a}$ give the total number of vaccinations with an mRNA vaccine in region $r$, on day $d$ in age-group $a$. Note, from Equation \eqref{eq:efficacy}, the basic parameters of this vaccine efficacy sub-model are the $\pi^{q,\textrm{mRNA}}_{t_k}$ and $\pi^{q,\textrm{AZ}}_{t_k}$. The expressions and parameters for the specification of $\alpha^q_{r,t_k,a}$ are analogous. The time dependence of both efficacies ($\pi$ and $\alpha$) for both vaccine types is through a piecewise constant specification with changepoints corresponding to the emergence of successive \covid\  variants each with increased ability to evade the vaccine-induced protection. Precise parameter values were based on published data in UKHSA vaccine surveillance reports \citep{UKHSA22}.

Figure \eqref{eqn:prob.infection} relates the force of infection, $\lambda_{r,t_k,a}$ to $b^{t_k}_{r,aa'}$, the per day probability of an infected individual of age $a'$ infecting a susceptible individual of age $a$ at time $t_k$. 
In the presence of a maximum of $Q$ doses of vaccine, $\lambda^{V_q}_{r,t_k,a}$ is the force of infection acting upon an individual who has received $q$ vaccine doses, is defined similarly to Equation (A.3) 
of the Online Appendix, with the assumption that the force of infection is reduced by a factor of $\pi^q_{r,t_k,a}$ in comparison to an unvaccinated individual:
    \begin{align*}
        \lambda^{V_q}_{r,t_k,a} &= 
        \left(1 - \pi^q_{r,t_k,a}\right) \lambda^{V_0}_{r,t_k,a}\\
        &=\left(1 - \pi^q_{r,t_k,a}\right) \left\{1 - \prod_{q'=0}^Q \prod_{a'=1}^{n_a} \left[\left(1 - b^{t_k}_{r,aa'}\right)^{\Istateprime{1}{q'}{r,t_k} + \Istateprime{2}{q'}{r,t_k}}\right]\right\}\\
        &= \left(1 - \pi^q_{r,t_k,a}\right) \left\{1 - \prod_{a'=1}^{n_a} \left[\left(1 - b^{t_k}_{r,aa'}\right)^{I^+_{r,t_k,a'}}\right] \right\}
    \end{align*}
    where $I^+_{r,t_k,a'} = \sum_{q',l}\Istateprime{l}{q'}{r,t_k}$, the sum of all infectious individuals in age group $a'$ in region $r$ at time $t_k$.

To account for the protection offered by the vaccination programme against severe illness, Equation \eqref{eq:nni} requires updating to reflect that fact that there is no longer a single homogeneous group of susceptible individuals
    \begin{align*}
        \nni_{r,t_k,a} &= \sum_{q=0}^Q S^{V_q}_{r,t_k,a}\lambda^{V_q}_{r,t_k,a}\delta t\\
        &=\sum_{q=0}^Q \left(1 - \pi^q_{t_k}\right) S^{V_q}_{r,t_k,a} \lambda^{V_0}_{r,t_k,a} \delta t\\
        &=\sum_{q=0}^Q \nni^{V_q}_{r,t_k,a}.
    \end{align*}
    To account for the impact of vaccination on severe disease, the convolution of Equation \eqref{eqn:convolutionA} is extended to give the number of severe events (e.g. death or hospital admission): 
    \begin{equation}\label{eqn:full.convolution}
        \mu_{r,t_k,a} = \sum_{l=0}^k f_{k - l} \sum_{q=0}^Q p^{V_q}_{r,t_l,a}\nni^{V_q}_{r, t_l, a}
    \end{equation}
    where $p^{V_q}_{r,t_l,a}$ is the infection-severity ratio describing the proportion of individuals who have had $q$ doses of vaccine who experience the severe event following infection. Let $\alpha^q_{r,t_l,a}$ describe the vaccine efficacy at time $t_l$ against a severe event (hospitalisation or death) conditional upon infection, we parameterise $p^{V_q}_{r,t_l,a} = (1 - \alpha^q_{r,t_l,a}) p^{V_0}_{t_l,a}$ for vaccination dose $q = 1, \ldots, Q$. Equation \eqref{eqn:full.convolution} simplifies:
    \begin{equation*}
        \mu_{r,t_k,a}
        = \sum_{l=0}^k f_{k - l} p^{V_0}_{t_l,a}\nni^*_{r, t_l, a}
    \end{equation*}
where, with $\alpha_0=0$,
\begin{equation}
    \nni^*_{r,t_l,a} = \sum_{q=0}^Q \left(1 - \alpha^q_{r,t_l,a}\right) \nni^{V_q}_{r,t_l,a}.\label{eqn:discount.infecs}
\end{equation}
The expression in \eqref{eqn:discount.infecs} is a `discounted' number of new infections, the effective number of infections that are unprotected by vaccination against the possibility of severe disease.

The actual fraction of infections that lead to a severe infection, $p^*_{r,t_k,a}$ is then derived from a weighted sum of the severity ratios for each of the levels of vaccination, weighted by the total number of infections in that vaccination strata
\begin{equation}\label{eq:actual.ifr}
    p^*_{r,t_k,a} = \frac{\sum_{q=0}^Q p^{V_q}_{r,t_k,a} \Delta^{V_q}_{r,t_k,a}}{\sum_{q=0}^Q \Delta^{V_q}_{r,t_k,a}}.
\end{equation}

\paragraph{December 2021 onwards -- waning immunity following emergence of Omicron variants}
    
With the emergence of Omicron it became necessary to consider reinfections and the model had to incorporate an element of waning immunity. In such a case it is needed to fully stratify by vaccination status, as can be seen in Figure \ref{fig:mod.figs}(C), where it is assumed that, even with prior infection, further vaccination will provide some boost to immunity. It is assumed that all individuals are equally likely to get vaccinated independently of infection status.

Waning of vaccination-acquired protection was simply accounted for through the piecewise-constant specification of the vaccine efficacy parameters. To account for the waning of immunity in those with a prior infection two new fully stratified model states are introduced, see Figure \ref{fig:mod.figs}(C). State W$^S$ contains individuals who are once again fully susceptible to infection with the currently circulating variant, having previously been infected. State W is an intermediate state between the `recovered' R states and the susceptible W$^S$ introduced to ensure that the duration of infection is not exponential-like and that those infected longer ago are those more likely to lose their protection. The expected duration of waning (time spent in R$^-$ and W combined) is $d_w$. The SIREN cohort study of UK healthcare workers
estimated that SARS-CoV-2 infection gave 85\% protection against reinfection over 6 months \citep{Hall2021}, consistent with a mean duration of $d_w = 534$ days, prior to emergence of the Omicron variant. Due to the significant immunity escape of the Omicron variant, it is necessary to `fast-track' individuals from the R$^-$ and $W$ states, so this expected duration is set temporarily to $d_w = 5$ days for five days. Following this readjustment, the mean duration of infection-derived protection from reinfection is $d_w = 117$ days, consistent with the 19\% protection over six months estimated in \citep{Ferguson2021}.

The full system of dynamic equations describing the model dynamics is expressed in system of equations 
(B.1) in Section B 
of the Online Appendix.

\subsubsection{Reproduction Numbers}\label{sec:Re}

The effective reproduction number, $\mathcal{R}_{t_k}$, quantifies how rapidly the pandemic is growing or declining at any given time $t_k$. Its expression, as used in \citet{BirBVGD21}, is given in Equations  (A.7)--(A.8) 
of the Online Appendix. In moving to the model represented in Figure \ref{fig:mod.figs}(B) and (C), the comparable expression is more complex now that the susceptible population has differential risk of infection within a single region and age group, due to differing levels of vaccination. However, $\mathcal{R}_{t_k}$ is still derived using the dominant eigenvalue, $\tilde{\R}_{t_k}$, of a next-generation matrix (NGM), $\tilde{\vec{\Lambda}}_{t_k}$. The \myth{(a,b)} entry of this matrix, in region $r$, is:
\begin{equation}\label{eqn:Rtilde}
  \tilde{\Lambda}_{r,t_k,ab} = \beta_{r,t_k} \tilde S_{r,t_k,a} \tilde C^{k}_{r,ab} d_I.
\end{equation} 
with $\tilde S_{r,t_k,a}$ being a weighted sum of all the susceptible individuals, where the weights are dependent on levels of vaccine-induced protection against infection, \textit{i.e.}
\begin{equation*}
\tilde S_{r,t_k,a} = \sum_{q=0}^Q \left(1 - \pi^q_{r,t_k,a}\right)\left(\Sstate{q}{r,t_k} + \Wstate{S}{q}{r,t_k}\right).
\end{equation*}
The time-$t_k$ reproduction number is then
\begin{equation}\label{eqn:Re}
\R_{r,t_k} = \R_{r,t_0} \frac{\tilde{\R}_{r,t_k}}{\R^*_{r,t_0}},
\end{equation}
where all other quantities are as described in Equation (A.7) 
of the Web Appendix.

There are three components of the reproduction number that evolve over time:  contact patterns, transmission potential $\beta_{t_k}$ and size of the susceptible population. To help understand the changing contribution of each of these factors to the overall reproduction number, we define a number of related quantities: (i) $\R^a$, the age-specific reproduction number, the average number of secondary infections caused by a single primary infection in age group $a$,
\begin{equation*}
\R^a_{r,t_k} = \frac{\R_{r,0}}{\R^*_{r,0}}d_I \beta_{r,t_k}\sum_{a'=1}^A \tilde S_{t_k,a'}\tilde{C}^{t_k}_{r,aa'} = \frac{\R_{r,0}}{\R^*_{r,0}}d_I \beta_{r,t_k}\left(\tilde{\vec{C}}^{t_k}_{r} \tilde {\vec{S}}_{t_k}\right)_a.
\end{equation*}
Instead of calculating a dominant eigenvalue of the NGM to average across age-groups, all that is needed is the matrix-vector product of the contact matrix and the susceptible population; (ii) $\R^W$, the reproduction number if the population was to somehow remain fully susceptible to infection, defined in \citet{Pellis2022} as the control reproduction number. This is derived in the same way as $\R$, except to replace $\tilde S_{r,t_k,a}$ in Equation \eqref{eqn:Rtilde} with $N_{r,a}$; (iii) $\R^B$, the reproduction number assuming both a fully susceptible population and no changes to the contact matrices. So we now seek to find the dominant eigenvalue of the matrix with elements
\begin{equation*}
\beta_{r,t_k} N_{r,a} \tilde C^{t_0}_{r,aa'} d_I,
\end{equation*}
so that any time trend in $\R^B$ can only be reflective of changes to the $\beta_{r,t_k}$.

\subsection{Bayesian Inference}

Continuing from \citet{BirBVGD21}, model parameters are estimated within a Bayesian framework. The expression of the log-likelihood takes the log-likelihood from \citet{BirBVGD21}, and adds a term due to the inclusion of the CIS prevalence estimates.

\subsubsection{Likelihood}\label{sec:melding}

The prevalence estimates are of log-counts of PCR-positive individuals, stratified by region and age-group, $\tilde Z_{r,t_k,a}$. These estimates are in the form of posterior means generated by the model described in detail in Section C 
of the Online Appendix and come with an attendant posterior standard deviation, $\tilde \xi_{r,t_k,a}$. These estimates come from independent region-specific models, but have a very strong auto-correlation over time. Therefore, to calculate the likelihood, estimates are only used every 14 days, for each age-group and region. A Bayesian melding approach is used to derive the likelihood \citep{Goudie2019}, where the estimates are treated as normally-distributed data, observed subject to the standard deviation attached to the estimate. If we denote the modelled log-prevalence to be:
\begin{equation*}
    \nu_{t_k,a} = \log\left(\sum_{d=0}^3 \left(\Istate{1}{d}{k} + \Istate{2}{d}{k} + \Rstate{+}{d}{k}\right)\right),
\end{equation*}
then
\begin{equation*}
    \tilde Z_{t_k,a} \sim \mathrm{N}\left(\nu_{t_k,a},\tilde \xi^2_{t_k,a}\right).
\end{equation*}

\subsubsection{Priors}\label{sec:priors}
Table \ref{tbl:pars} gives an overview of all the parameters that are treated as unknown and gives an overview of the prior distributions used and the source of information on which the priors are based, where applicable. Note that parameters are partitioned into two categories according to whether they apply `globally' to all regions, or are regionally stratified.

\begin{table}[!ht]
\centering
\caption{Model parameters with assumed prior distributions or fixed values}\label{tbl:pars}
  \begin{tabular}{p{0.05cm}p{5.3cm}p{5.3cm}}
    \textbf{Name}&&\textbf{Prior source}\\
    \hline
    \hline
    \multicolumn{2}{l}{\textbf{Regional parameters}}&\\
    &Contact matrix modifiers, $m_{r,l}$&Log-normal priors based on analysis of \citet{HouRPPBEJEW21}\\
    &Exponential growth, $\psi_r$&$\Gamma(31.36, 224)$, derived through mapping from a flat distribution over $R_0$ given sampled values of $d_I$ and assumed value of $d_L$.\\
    &Initial infection, $I_{0,r}$&Uninformative, see \citet{BirPCZD17}.\\
    &Time-variation in transmission potential, $\beta_{t_k,r}$&Evolves according to a region-specific log-random walk, with variance $\sigma^2_{\beta}$.\\
&&\\
    \multicolumn{2}{l}{\textbf{Global parameters}}&\\
    &Mean infectious period, $d_I$&2 + $\Gamma(1.43, 0.549)$, based on \citet{LiGWWF20}.\\
    &Residual duration of PCR-positivity, $d_R$&1 + $\Gamma(32.2,2.6)$, based on \citet{CevTLMSH20}.\\
    &Infection-hospitalisation rate $p_a$&Initial, `wild-type' estimates based on \citet{Verity2020}. Temporal changes in severity all assumed uninformative with zero mean.\\
    &Negative Binomial over-dispersion, $\eta$&Uninformative $\Gamma(1, 0.2)$\\
    &Step-size on log-scale of weekly variation in transmission, $\sigma_{\beta}$&Informative $\Gamma(1, 100)$.\\
    &Serological test sensitivity, $\ksens$&Based on convalescent sera, $\beta(52.9,17.9)$ (EuroImmun) and $\beta(457,13.2)$ (Roche-N).\\
    &Serological test specificity, $\kspec$&Based on pre-COVID-19 sera, $\beta(314,3.18)$ (EuroImmun) and $\beta(672, 1.35)$ (Roche-N).\\
    \hline
    \end{tabular}
\end{table}

\subsection{Dealing with computational complexity}

By March \myst{31} 2023, the estimation process involved exploring a joint posterior over 692 parameters. The naive MCMC algorithm used in \citet{BirBVGD21} proved insufficiently powerful to achieve convergence within a reasonable time-frame. Instead we extend to using an adapted version of the Adaptive Metropolis with Global Scaling (AMGS) algorithm \citep{Andrieu2008}. Further detail on this algorithm can be found in the Online Appendix, Section 
D. To implement this algorithm, the `global' parameters were updated as a single block update using AMGS, followed by the updating of $n_r$ regional parameters blocks in parallel.

\section{Results}\label{sec:results}

To the end of March 2023, analysing the pandemic over 1139 days of data, we are able to provide here a detailed nowcast, providing: an estimate of the `current' state of the epidemic; and a full reconstruction of the epidemic history, illustrating its development over time, deconstructing the history of transmission and quantifying the impacts of the vaccination programme.

\subsection{Nowcast, March 2023}\label{sec:march23}

\begin{figure}
\centering
\includegraphics[width=\linewidth]{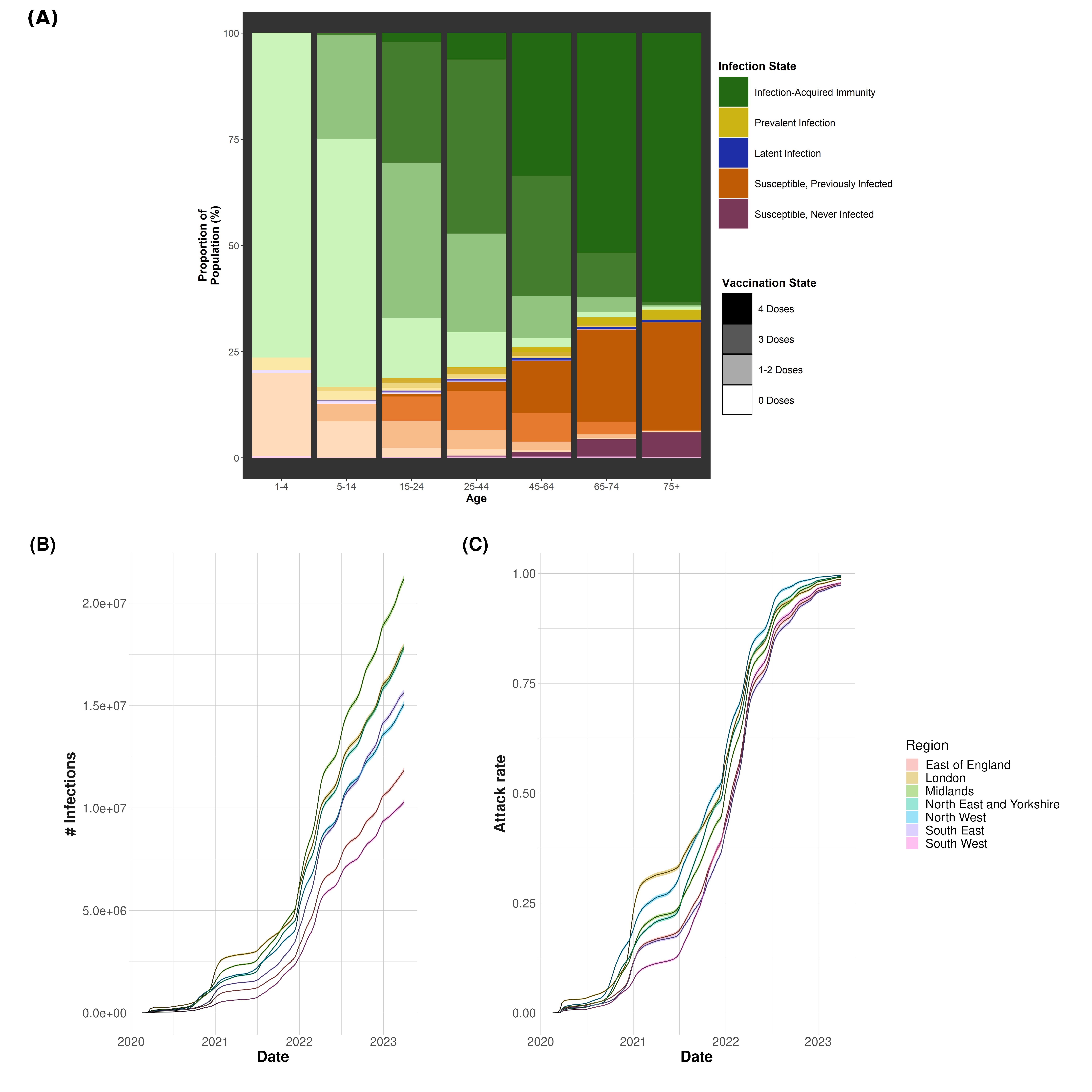}
\caption{(A) Proportion of the population by infection status (susceptible, exposed/latent, infectious, infection-acquired immunity) and number of vaccine doses, stratified by age. The maroon colour palette indicates the fraction of the population who are susceptible and never infected, the orange palette, those that are in susceptible states having had a previous infection, the blue and the olive green palettes indicate latent and prevalent infections respectively and the green bars represent those currently with infection-acquired immunity. (B) and (C) present the regional cumulative infections and attack rate (fraction of the population ever infected) over the course of the pandemic respectively.}\label{fig:snapshot}
\end{figure}

Figure \ref{fig:snapshot}(A) gives a snapshot of the pandemic on \myst{31} March 2023, showing the estimated age-specific distribution of the population over the disease and vaccination states (see Supporting Information, Figure (E.1) 
for a regional breakdown). 
The maroon fraction represents those susceptible having never previously been infected: this increases with age, ranging from 0.0076\% (95\% credible interval (CrI), 0.0060\%-- 0.0097\%) in the 5--14 age group, the majority of whom have never been vaccinated, to the two age-groups over the age of 65 where 4.0\%--6.2\% have never been infected, almost all of whom have had at least three vaccine doses. The orange represents the proportion who are susceptible having been previously infected. There is much less heterogeneity in this `susceptible, previously infected' proportion across age groups, from 25.9\% (25.3\%--26.4\%) in the 75+ to 12.6\% (12.0\%--13.3\%) in the 5--14. Infection-acquired immunity (in green) is greatest in the 5--14, estimated at 83.4\% (83.0\%, 83.8\%). In older age-groups, a higher fraction of the infection-acquired immunity (green) group have received two or fewer vaccine doses than in the susceptible never-infected (maroon) fraction, highlighting the protective effects of the vaccine campaign against infection.
  
\begin{table}

\caption{\label{tab:incidence}Estimates of cumulative infection, attack rate up to \myst{31} March 2023 by region}
\centering
\begin{tabular}[t]{lll}
\toprule
Region & Cumulative Infections, $\times 10^6$ & Attack rate\\
\midrule
England & 110 (109--110) & 98.6\%(98.5\%--98.7\%)\\
\midrule
East of England & 11.8 (11.7--12.0) & 97.8\%(97.5\%--98.0\%)\\
London & 17.8 (17.7--18.0) & 98.7\%(98.5\%--98.8\%)\\
Midlands & 21.2 (21.0--21.4) & 99.1\%(99.0\%--99.3\%)\\
North East and Yorkshire & 17.8 (17.6--18.0) & 99.3\%(99.2\%--99.5\%)\\
North West & 15.1 (14.9--15.3) & 99.6\%(99.5\%--99.7\%)\\
South East & 15.6 (15.5--15.8) & 97.3\%(97.1\%--97.6\%)\\
South West & 10.3 (10.2--10.4) & 97.9\%(97.7\%--98.1\%)\\
\bottomrule
\end{tabular}
\end{table}

Table \ref{tab:incidence} displays estimates of the cumulative infection over the course of the pandemic, by region. Due to waning immunity leading to reinfection, the cumulative number is substantially larger than the attack rate (the fraction of a population to ever have been infected) would suggest: an estimated 98.6\%
of the population has acquired 109.6M
  infections. The attack rate is highest in the North West and the North East and Yorkshire, whilst it is lowest in the South East, consistent with the regional snapshot figure (see Supplementary Information, Figure (E.1).  
Figure \ref{fig:snapshot}(B) and (C) show, respectively, the estimated cumulative infections and attack rates, both over time and by region. It is clear that the cumulative infections increase rapidly from December 2021 at least until April 2022, whereas the growth in attack rates is much more gradual.
On December \myst{1}, 2021, we estimate a cumulative 25.8M (25.6M--25.9M)infections. Assuming this date to be the beginning of the Omicron waves, we estimate the Omicron variants have been responsible for 76.5\% (76.4\%--76.6\%)of the total number of infections over the course of the pandemic.

\begin{table}

\caption{\label{tab:firsttimers}Estimates of the fraction of infections that are re-infections}
\centering
\begin{tabular}[t]{ll}
\toprule
Date & Proportion of re-infections\\
\midrule
2020-03-23 & 0.0\% ( 0.0\%-- 0.0\%)\\
2021-01-04 & 0.9\% ( 0.9\%-- 0.9\%)\\
2021-10-11 & 10.0\% ( 9.9\%--10.2\%)\\
2022-01-03 & 44.7\% (44.3\%--45.0\%)\\
2022-03-14 & 45.1\% (44.8\%--45.4\%)\\
2022-07-04 & 57.7\% (57.4\%--58.1\%)\\
2023-03-31 & 96.0\% (95.5\%--96.5\%)\\
\bottomrule
\end{tabular}
\end{table}

Overall,
49.7\% (49.4\% -- 49.9\%)
of all infections are estimated to be re-infections. 
Table \ref{tab:firsttimers} shows how these reinfections are distributed over the course of the pandemic. Dates are chosen to represent the dates on which the estimated number of infections peaked, corresponding to the wild-type, Alpha, Delta and Omicron BA.1, BA.2 and BA.4/5 variants. Though the proportion increases between peaks, there is a plateau in the proportion of re-infections between January and March 2022. 

\subsection{Epidemic Reconstruction}

\paragraph{Severity}

\begin{figure}[!ht]
\centering
\includegraphics[width=\textwidth]{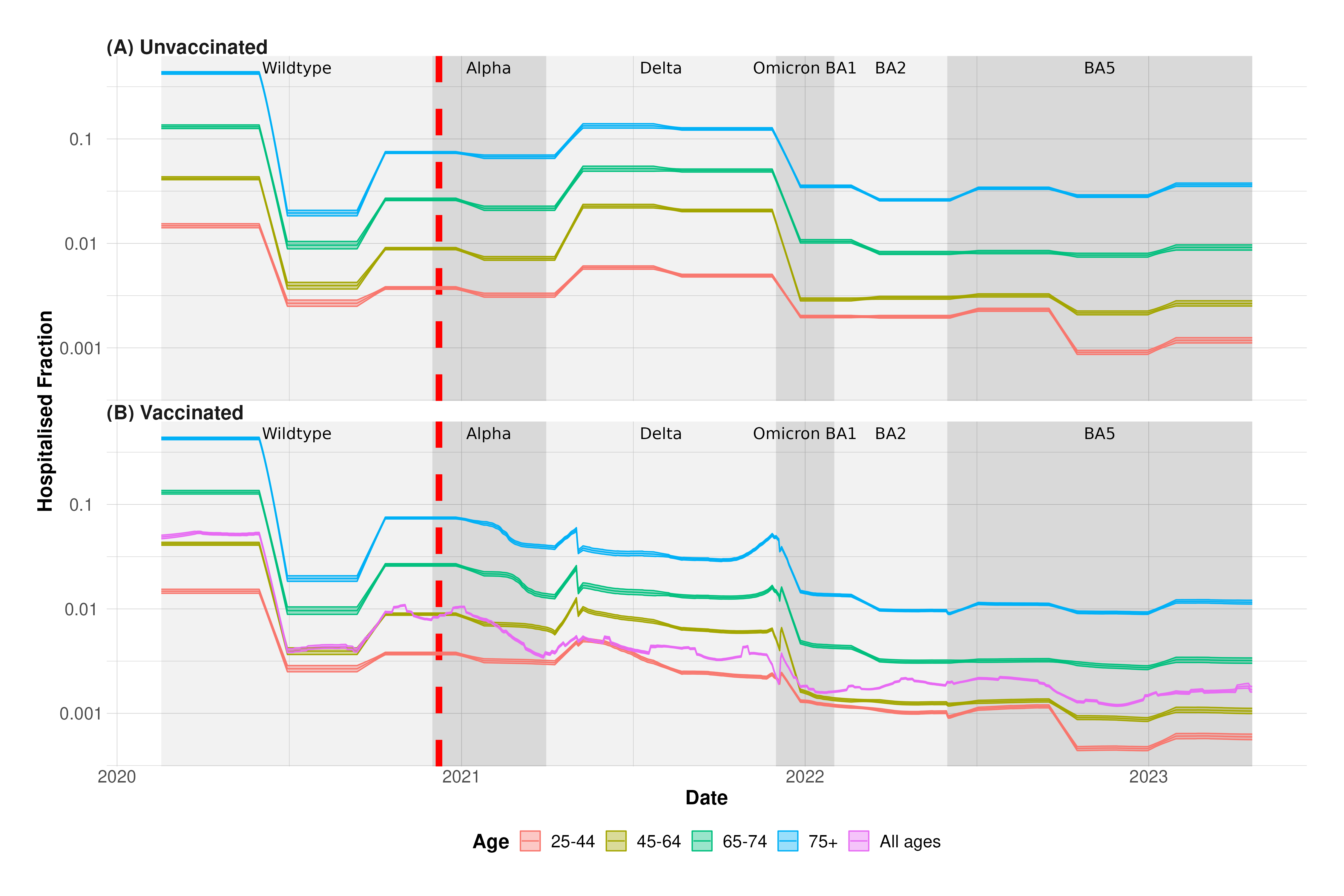}
\caption{Posterior summaries of the IHR (panel A) and pIHR (B) over time for the 25--44, 45--64, 65--74 and 75+ age groups together with an age-averaged quantity in (B). Estimates for under-25s are omitted for clarity, but they have the same temporal trends as the 25--44 (by construction) but with different initial base rates of survival.}\label{fig:ifr}
\end{figure}

Estimates of severity are shown in Figure \ref{fig:ifr}, which displays the log of the infection-hospitalisation ratio (IHR) by age and time. This is the proportion of infections on a given day that are subsequently hospitalised due to the severity of their symptoms. In Fig \ref{fig:ifr}(A) we plot posterior summaries of the piecewise-constant IHR parameters, $p^{V_0}_{r,t_k,a}$, the IHR among unvaccinated individuals. This shows the relative severity of infection in each variant-defined era of the pandemic, though the Omicron variants may appear to have lower severity due to a large fraction of these infections being reinfections. The initially high severity of the wild type virus decreases in summer 2020 before increasing again in the Autumn towards its earlier high level in all age groups. There is no real change in severity with the emergence of the Alpha variant, but there is a significant increase in the severity of the Delta variant, which persists until the milder Omicron variant emerges in December 2021.

In Fig \ref{fig:ifr}(B) we present `population IHRs' (pIHRs): the actual fraction of individuals who are admitted to hospital, $p^*_{r,t_k,a}$ (see Equation \eqref{eq:actual.ifr}) which is heavily influenced by vaccination uptake and efficacy and the consequences of previous infection. As in Fig \ref{fig:ifr}(A), there is a clear pattern across the age groups with the older individuals having the highest severity and orders of magnitude differences between age groups, though the IHRs in the 25--44 and 45--64 become similar throughout 2022. Commensurate with the staggered roll-out of the vaccination programme, we can see that throughout 2021 there is a gradual decline in all pIHR estimates. This is initially evident in the over--75s, before sequentially moving to younger age groups, despite the emergence of the more severe Delta variant \citep{Twohig2021}.
The decreases flatten off towards late 2021, before a major fall in the pIHR due to the emergence of Omicron. Also in (B) we calculate an age-averaged pIHR using infection numbers as weights. The age-averaged pIHR diverges from the age-specific pIHRs at various times. This is due to shifts in the age profile of new infections: as the proportion of infections in older age groups increases, the pIHR will increase, and vice versa.


\paragraph{Transmission}


The effective reproduction number $\R_{r,t_k}$ is seen from Equation \eqref{eqn:Re}, and surrounding text, to be a function of three quantities that vary over time: population susceptibility; contact rates; and changes in transmissibility quantified through the time-varying parameter $\beta_{t_k,r}$.

\begin{figure}[!ht]
  \centering
\includegraphics[width=\textwidth]{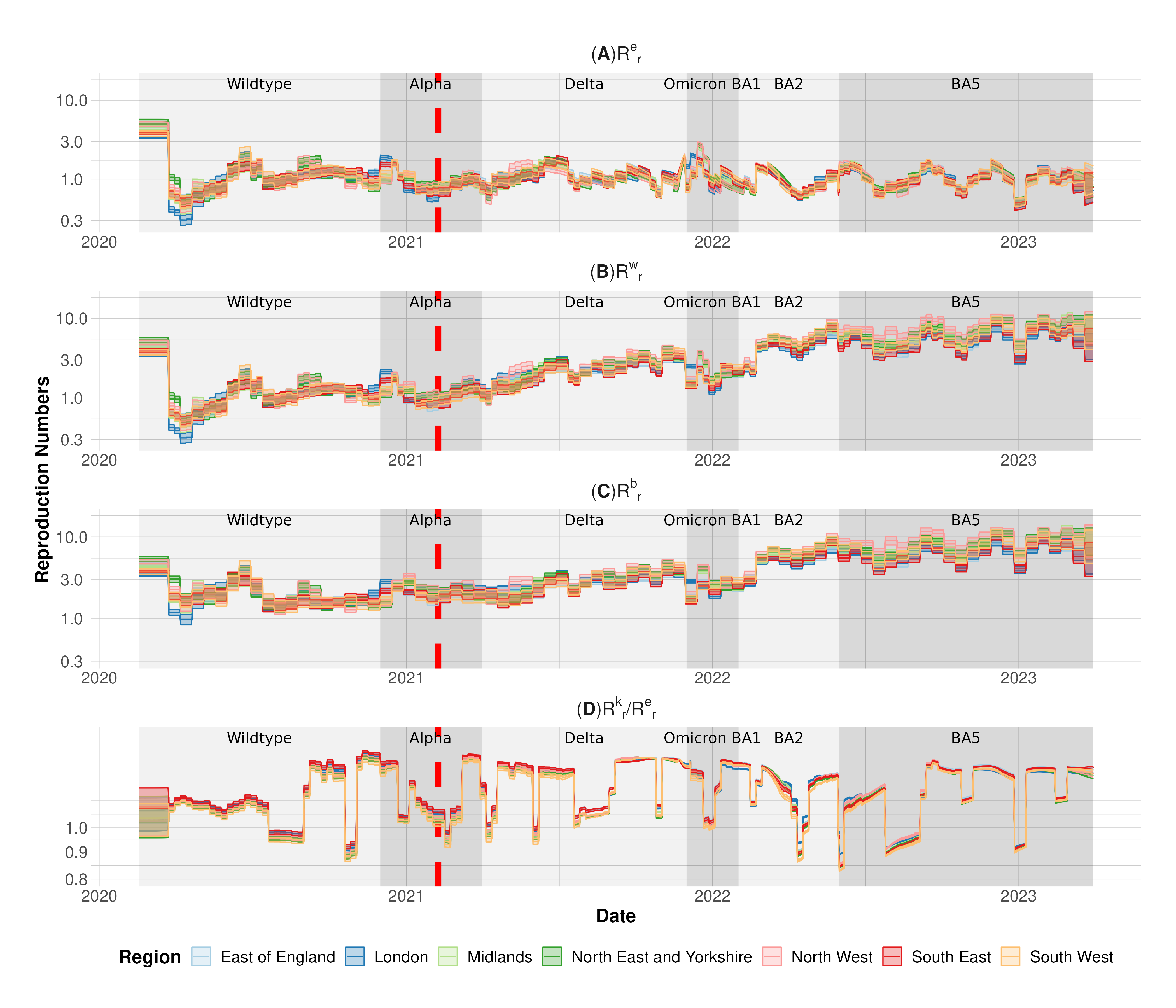}\\
\caption{(A) Estimated effective reproduction number, $\R_{r,t_k}$, by region; (B) Estimated reproduction number, $\R^W_{r,t_k}$ in the absence of susceptible depletion; (C) Estimated reproduction number $\R^B_{r,t_k}$ with contact rates assumed constant over time; (D) $\R^3_{r,t_k} / \R_{r,t_k}$, where $\R^3_{r,t_k}$ is the estimated reproduction number for children $<15$ years.}\label{fig:Rs}
\end{figure}

Figure \ref{fig:Rs}(A) shows the evolution of the effective reproduction numbers $\R_{r,t_k}$ over time.
It is estimated that the March 2020 lockdown induced a huge fall in $\R_{r,t_k}$ in all regions, from values in the range 3.4 to 5.4 down to values in the range 0.38 to 0.96 in early April, with the steepest decline in London, and the most gradual in the North East and Yorkshire. Aside from this, the regions appear homogeneous except around the emergence of new variants. We estimate a high value of $\R_{r,t_k}$ for London and the South East before the end of 2020 (due to the emergence of the Alpha variant in Kent), before then falling most sharply in the new year. In April-May 2021, around the time of the emergence of Delta, the North West leads the way (in line with the emergence of Delta outbreak being localised here \citep{Tor21}). London has the highest $\R_{r,t_k}$ in the early stages of the Omicron era, before dropping below that of other regions in late December when the North East and Yorkshire have the highest transmission rates. The second and third Omicron waves appear more homogeneous across regions.
Note also that
the behaviour of the plotted curves display a gradual decline between sudden changes due to the structural changepoints in the value of $\beta$ and changes in the contact matrix (see Methods section \ref{sec:Re}). These within-week declines are due to the depletion of the susceptible population and these are steeper at times of higher incidence (and before the high rates of waning immunity in the Omicron era).

In 
Figs \ref{fig:Rs}(B)-(D) we dissect the reproduction number to understand a more about what is driving transmission in each of the regions. Figures \ref{fig:Rs}(B) and (C) present estimates of $\R^W_{r,t_k}$ and $\R^B_{r,t_k}$, the effective reproduction number if there was no waning of immunity, and if there was no waning of immunity or changes in the contact matrix respectively (see Methods section \ref{sec:Re}).
The effect of removing the waning in (A) exaggerates the increase in $\R^W_{r,t_k}$ in London towards the end of 2020 and singles out increased transmission in the North West over the second half of 2022. Clearly, as this high transmission does not feature in the plots of $\R_{r,t_k}$, it is being suppressed due to diminished population susceptibility.

It is interesting to note that the reproduction number $\R^W_{r,t_k}$ (and $\R^B_{r,t_k}$) drops around the emergence of Omicron. This suggests that the increasing incidence (and hence $\R_{r,t_k}$, see Figure \ref{fig:Rs}(A)) around this time is being driven by boosted susceptibility (diminshed immunity to the new variant) rather than an increased transmissibility of the virus. The rate of transfer of immune individuals in the model to the susceptible states is an assumed quantity and a lower value could lessen this drop in $\R^W_{r,t_k}$. Subsequent omicron waves, however, do show a gradually increasing $\R^W_{r,t_k}$ suggesting that immunity escape cannot explain these waves alone.

In Figure \ref{fig:Rs}(C), where we have removed the effects of changing contact patterns, the drop in $\R^B_{r,t_k}$ following the first lockdown is now only marginal in comparison to $\R^W_{r,t_k}$, highlighting the degree to which lockdown measures limiting movement and interaction suppressed transmission at this time. $\R^B_{r,t_k}$ is now only a function of the $\beta_{r,t_k}$ so Figure \ref{fig:Rs}(C) effectively shows only the changes to the underlying transmissibility due to the particular variant mix at time $t_k$ combined with other extrinsic factors. By March 2023, this is taking values in the range 6.2--14 for the North West, an increase from the initial reproduction number by a factor in the range (1.1--2.8).
  
The fourth variation on the reproduction number that we consider is $\R^3_{r,t_k}$, the number of secondary infections caused by a single primary infection in the 5--14 group. Figure \ref{fig:Rs}(D) plots the ratio $\R^3_{r,t_k}/\R_{r,t_k}$, measuring the relative contribution of the 5-14 age-group (the age group most closely aligned with school-aged children) to overall transmission rates. Prior to the initial lockdown on March 23, 2020, there is great heterogeneity in this quantity, but it soon plateaus until the long summer school holiday around July/August 2020. Throughout the timelines there are sudden periodic drops corresponding to school holidays and closures limiting children's capacity to transmit infection. Between times it can be seen that rates of transmission were elevated amongst children at the beginning of the second wave and at the start of the Alpha era until schools were closed at the end of December 2020. The increased contribution of children to transmission is also present throughout the Delta wave (when schools were not closed to mitigate transmission). Throughout the Omicron waves the ratio is not as large as during the Alpha and Delta waves, it is still (in most regions, for most of the time) higher than during the initial wild-type variant era. These elevated relative rates of transmission in children will be promoted in part by the protection offered to older age groups from the vaccination campaign. Only during Omicron do children have any protection through immunisation.


    

\paragraph{Consistency with the data}


\begin{figure}[!ht]
\centering
\includegraphics[height=0.7\textheight]{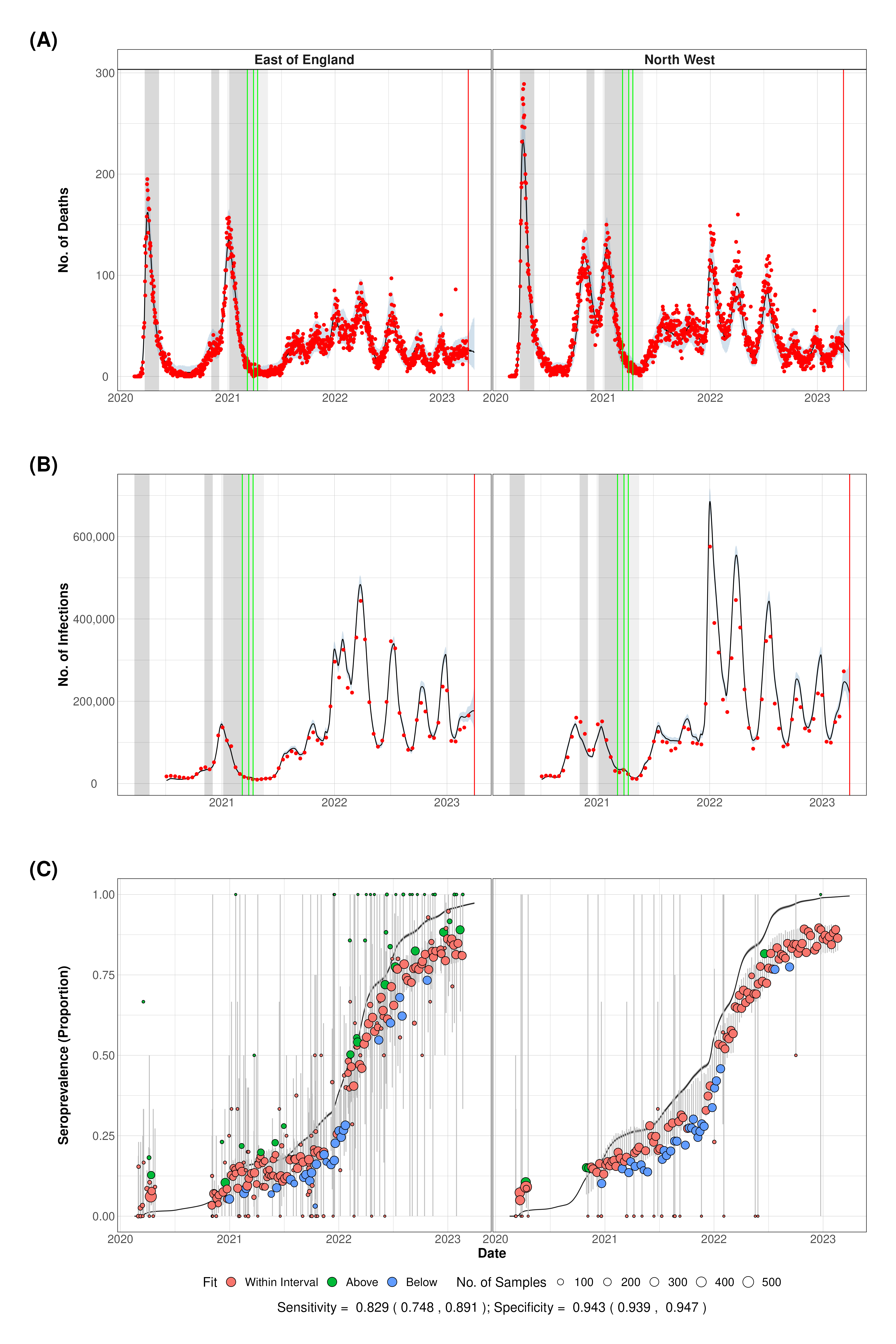}
\caption{Goodness-of-fit of the model to the three main types of data: admissions (A); prevalence (B); and serology (C). For each type of data the fit to the data is shown in two regions, the East of England and the North West.}\label{fig:gof.plots}
\end{figure}

Figure \ref{fig:gof.plots} show how faithful the model is to the data in two of the seven study regions, East of England and North West (see Supplementary Material Figures E.2 -- E.4 
for other regions). Figure \ref{fig:gof.plots}(A) shows a good fit to the admissions data, and (B) suggests similar performance in capturing the ONS prevalence estimates (\ref{fig:gof.plots}(B)).
The posterior mean duration of PCR positivity is 11.10 (11.03, 11.16) days, indicating that, like the admissions data, prevalence is a smoothed and lagged indicator of incidence.

Figure \ref{fig:gof.plots}(C) illustrates the fit to the seroprevalence data. In the figure, the black curves plot the posterior median (and 95\% CrI shaded grey) for the proportion of the population over-15 years who have ever been infected over time - the underlying seropositivity. However, the observed data typically lie below these lines due to the tested samples: (a) not being taken from the under-18s or over-70s as they are unable to donate blood; (b) do not constitute an unbiased sample from the population and may over-represent some age groups; and (c) being subject to an imperfect testing process which has a sensitivity and specificity, the estimates for which are specified in the figure. For these reasons, we would not expect the data points to lie along the grey lines.
In this figure, the plotted points record the observed seropositivity by day of sample (plus 25 days to allow for the development of an immune response). Red dots indicate points that lie within predictive credible intervals (the vertical bars plotted in pale grey), the green dots are data points that lie above the predictive intervals and the blue dots lies below the predictive intervals. A pattern can be detected here, blue points, signifying an over-estimation of the sero-positivity are (mostly) present throughout 2021, with the green points appearing less frequently and outside of this period. 
The prevalence of antibodies in the NHSBT data might be negatively biased due to blood donors constituting a biased sample of individuals more likely to get vaccinated. That the divergence occurs in 2021 largely covering the elongated Delta wave and immediately following the launch of the vaccination campaign supports this notion. The biases inherent in population prevalence data that did not account for vaccination status have been displayed elsewhere, e.g. \citet{Pouwels2023}.

\paragraph{Direct and Indirect Impacts of Vaccination}

With the vaccination efficacy parameters of the model set to zero, existing posterior samples were used to evaluate the model and simulate epidemics that would've occurred in the absence of an effective vaccine, under an assumption that the vaccination programme had no influence on pandemic policy or on behaviour. By comparing, for each posterior sample, the
difference between these counterfactual epidemic curves and the ones generated assuming 
effective vaccination, we can derive a
posterior distribution for the number of infections saved and for the
number of hospital admissions prevented. This was a routinely task reported in PHE/UKHSA
vaccination surveillance reports over the course of the first three
quarters of 2021 \citep[e.g.][]{UKHSA21}, stopping when it
became too unrealistic to assume that there would be no further government legislation and that the population would continue to behave in exactly the same way as they did in
the presence of an effective vaccination programme.

\begin{figure}
    \centering
    \includegraphics[width=\linewidth]{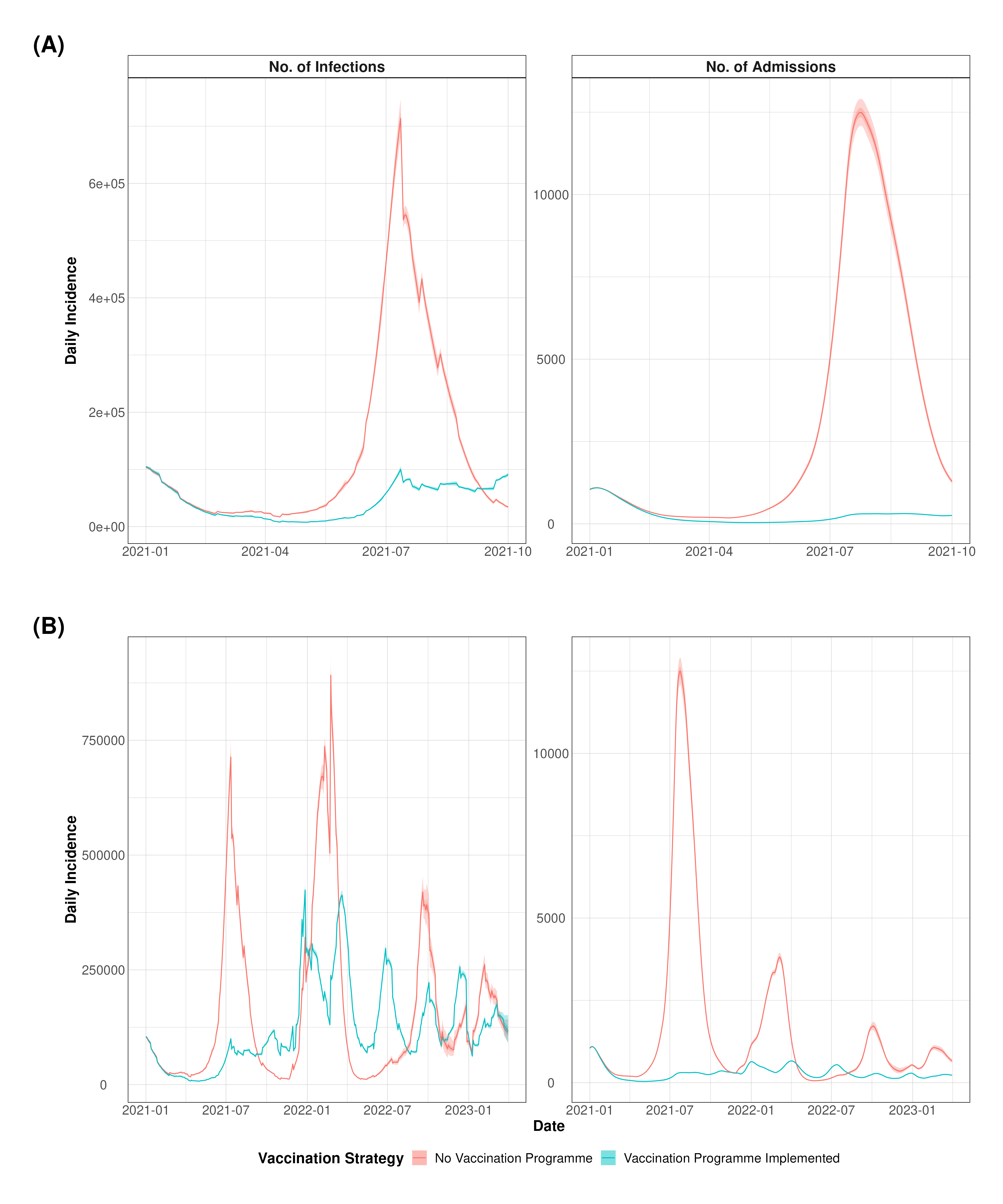}
    \caption{In the top row we have plots of the number of infections and deaths prevented as estimated on the \myth{17} September, 2021, and the bottom row correspond to the same analysis carried out on \myth{25} March, 2022}
    \label{fig:VLM}
\end{figure}

In Figure \ref{fig:VLM}, both the `true' epidemic curves are presented alongside the counterfactual scenario in terms of the number of infections (left column) and the number of admissions (right column). The top row looks at the period over which the vaccine surveillance reports were produced, whereas the bottom row looks at the entire vaccination era. By \myst{1} October, 2021, 25.5M (25.3M--25.7M) infections and 751K (735K--769K) admissions had been prevented. It is clear from the admissions plots in Figure \ref{fig:VLM} that there would have been a surge in hospital admissions in 2021 that health services would have been ill-prepared to handle, one that is far in excess of the winter 2020-21 surge that stretched services to a dangerous extent. Beyond October 2021, in the infections panel of Figure \ref{fig:VLM} one can see that the complex interplay between immunity, infection and waning means that there are further peaks in both the vaccination and no vaccination curves, with decreasing synchronicity. However, because of the enduring protection against severe disease that the vaccine appears to confer, this intersection of the two curves is barely seen in the admissions plot.
 


\paragraph{Deaths and Admissions}
This choice was made to switch from using data on deaths to the admissions dataset due to the withdrawal of free access to testing for all except hospital patients and to people living or working in `high-risk settings' from the \myst{1} April, 2022. This policy decision may have impacted the ability to ascertain all deaths due to SARS-CoV-2 infection while symptomatic and asymptomatic testing continued in hospitals beyond this time.
A look at how this choice of data time series impacts upon our estimated epidemics can be seen in Section E 
of the online appendix.

\section{Discussion}

We have presented here an in-depth examination of one of the key models contributing to the evidence base underpinning the pandemic response in England.
Assimilating data from a variety of sources, this model was implemented weekly for over three years, and we have presented a reconstruction of the pandemic dynamics over this period, uncovering fluctuations in transmission and quantifying the impacts of lock-down measures and the vaccination programme. 

As can be seen in  Section \ref{sec:results}, the model structure permits estimation of a range of `nowcast' quantities beyond reproduction numbers. In particular we have shown how a snapshot of the susceptibility profile of the population can be obtained. By March 2023, this shows that almost all of the population have had a previous infection. Due to the Omicron variants, most infections were re-infections by mid-2022 and the population averaged almost two infections per person.
Reproduction numbers themselves have been deconstructed to have a look at the individual roles of movement, immunity and transmissibility and how these have changed over time. The Omicron variants seemingly spread so effectively due to evasion of prior immunity. Additionally, we can quantify the impacts of pandemic mitigation: quantifying the reduction in transmission due to the March 2020 lockdown; and the impact of the vaccination programme on both the case-severity ratio and the total healthcare burden placed on the country and its healthcare services.

These insights, are, however, conditional on a number of untested model assumptions. In particular, we have accommodated the impacts of vaccination on susceptibility to infection and severe illness, but not on transmissibility. Higher Ct values (i.e. lower viral loads) are typically detected in prevalent infections, and this may lead to reduced transmissibility \citep{Eyre2022}. The arrival of the Omicron variant required a sudden shift of individuals from immune to susceptible model compartments in an ad hoc manner and this influences the conclusions drawn from \ref{fig:Rs}.

Also, a natural question for an evidence synthesis model of this type is to understand the role of the different data sources. We illustrate this through a sensitivity analysis to the inclusion of the CIS, the unprecedented large-scale survey that run from April 2020 to the end of March 2023. 

\subsection{Sensitivity to ONS CIS inclusion and timeliness}\label{sec:disc2}

To do this, we look at results from three points in time during the pandemic, at the end of March in 2021, 2022 and 2023. For each of the three times of analysis, inference was drawn from the model using three different levels of prevalence data: `full' where the prevalence estimates are included in full; `minus8', where the last 8 weeks of prevalence estimates are excluded; and `none' where the prevalence estimates are excluded completely. In each case, the model was implemented `as was', i.e. using the suite of data and prior information that were available at the time of the original analysis. The most significant consequence of this is that the 2021 analyses were based on data on deaths rather than hospital admissions, and only used the early serological data obtained using the EuroImmun assay. It was not anticipated that these inter-year differences would introduce any systemic change in the estimated infection curve derived from the `full' 2023 analysis, taken here as a gold standard.

In particular, to quantify the differences between the analyses, we look at the estimation of the size and the timing of peaks in infection corresponding to different SARS-CoV-2 variants: wild-type, Alpha, Beta, Omicron BA1 and Omicron BA2.
For all analyses, the peaks are defined to be the highest estimated incidence within two weeks of the peaks as estimated in the gold-standard.
In practice, the majority of the estimated peaks were temporally co-incident, with the 2021 analysis having estimated peaks for the Alpha-wave occurring 1-2 weeks earlier. 
The Delta wave consisted of two peaks, here we consider the earlier July peak as it is more temporally remote from other waves. As the 2022 analyses took place less than two weeks after the Omicron BA2 peak, there is greater uncertainty in the timing of this peak and it could have occurred across a range of dates.

\begin{figure}
\centering
\includegraphics[width=\linewidth]{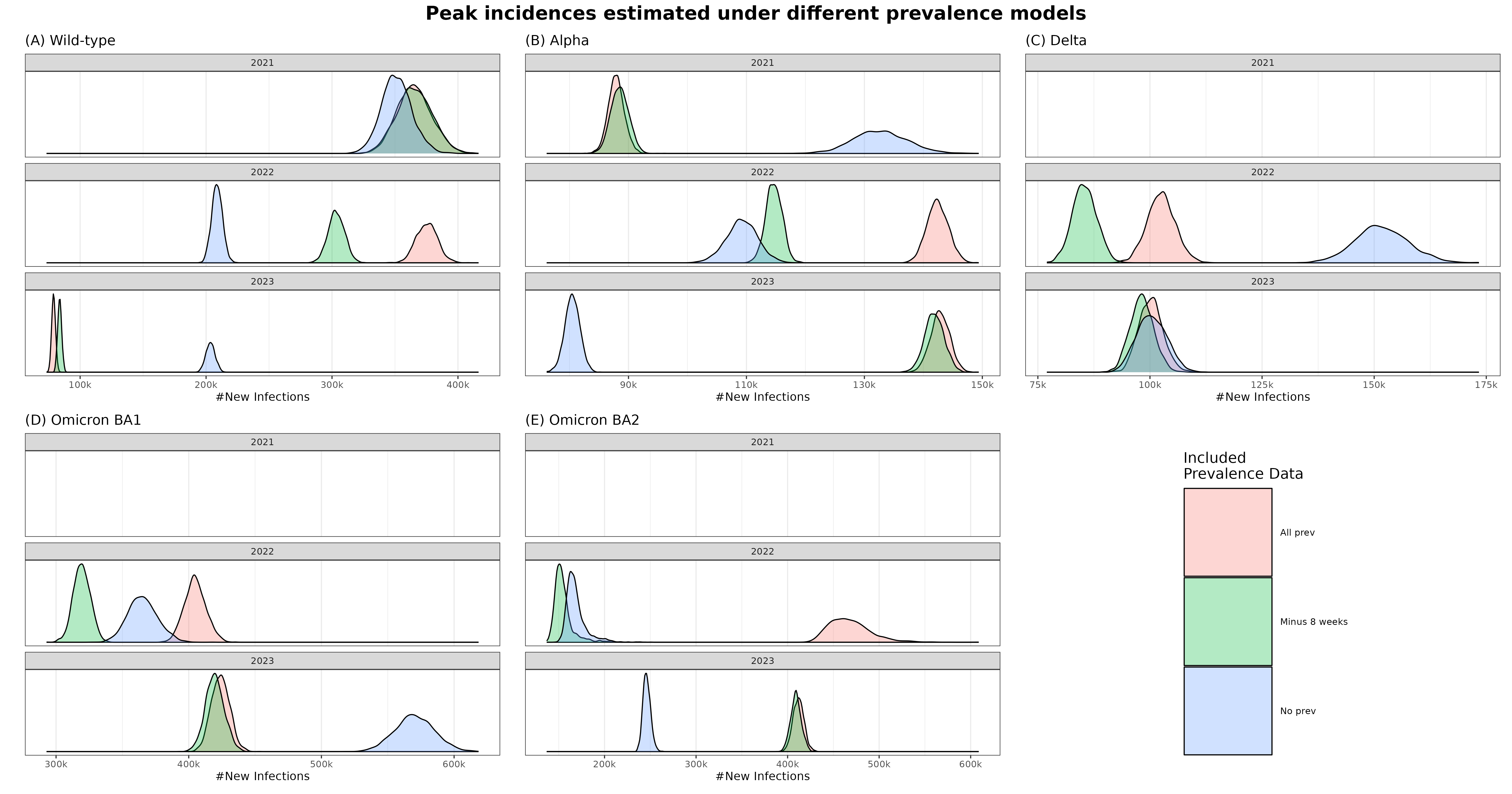}
\caption{Kernel density estimates of the sizes of the wild-type, Alpha, Delta and Omicron BA1 and BA2 peaks by the year in which the estimate was made and the level of prevalence data included.}\label{fig:peak.dens}
\end{figure}

Figure \ref{fig:peak.dens} shows how the estimated sizes of the various peaks vary with time and with the different levels of prevalence data included in the analysis. The wild-type peak, induced by the lockdown, is very inconsistently estimated, and is estimated to be smaller as data progressively accumulate. This wild-type peak is the only peak period not informed by the CIS estimates and it is the only peak at which there is a major discrepancy between the 2022 and 2023 analyses when all the prevalence data are included. In general, in 2023, the loss of eight weeks of prevalence information appears to have very little impact, unsurprising given the temporal separation between the analysis time and the time of the peaks. However, when no prevalence is included, the peaks look very different and vary considerably between years: inclusion of the prevalence data is necessary for the stable estimation of the peaks.

Further evidence of the utility of the prevalence data comes from the 2022 estimation of the timing of the Omicron BA2 peak (Figure \ref{fig:peak.dens}E). This estimated peak is unique, in that it is the only example where the peak had yet to be observed in either the prevalence or admissions datasets at the time of analysis.
This has an implication on the estimation of the timing of the peak in hospital admissions, which is also more precisely identified through the inclusion of prevalence information:
the gold-standard analysis produced an estimate of the peak in hospital admissions on 01/04/22 (posterior mode), whereas the comparable 2022 analysis estimated the peak on 07/04/22, six days later. Removing eight weeks of prevalence data pushes this estimated peak further back to 09/04/22, and with no prevalence at all the estimated peak is even later, on 11/04/22.

This analysis represents an initial attempt to understand how the different datasets in such an evidence synthesis contribute to the identification of different features of the overall model and influence the key outputs that underscore pandemic policy. More in-depth studies of this type are required to enable more informed choices of parameters values that can be estimated on the basis of the available data. Such a study could also extend to looking at mis-specification of the underlying MRP model used to produce the prevalence estimates. For example, recent work \citep{Pouwels2023} has quantified the sensitivity of prevalence estimates to additional stratification of the CIS study population by vaccination status.
Despite the differences in the results obtained when using differing levels of prevalence data, it still proved possible to achieve a very good fit to each of the datasets (see Online Appendix for detailed plots for each data stream).

It is imperative that, despite the successes of pandemic real-time modelling, model development continues in the inter-pandemic period, to lessen the reliance on some of the pragmatic assumptions discussed above, as well as to incorporate demographic processes and increase spatial resolution and temporal variation in parameters. The corresponding increase in model complexity and the need to incorporate a greater range of information sources will heighten the difficulty of delivering pandemic inference in a timely manner to policymakers. This motivates the continual development and application of state-of-the-art Bayesian methods for online inference to pandemic modelling. Furthermore, it will be crucial to better understand the value of different datasets to both nowcasting, model projections and the identifiability of key epidemic parameters so that this process of model development maximises the utility of the pandemic data streams likely to be available.

\section{Acknowledgements}

For the purpose of open access, the author has applied a Creative Commons Attribution (CC BY) licence to any Author Accepted Manuscript version arising from this submission.
PJB, EvL, TF, NG, AC, are all wholly or in-part funded by UKHSA, PJB, AA, JK and DDA are funded by the UK Medical Research Council programme MRC\_MC\_UU\_00002/11. DDA and AC are additionally part-funded through the NIHR Health Protection Research Unit in Behavioural Science and Evaluation at University of Bristol, in partnership with UKHSA. JB received support by the EPSRC (EP/R01856/1). Prior to the pandemic, this project was developed under a grant from the National Institute for Health Research (HTA Project: 11/46/03). We gratefully acknowledge the access to the data from the United Kingdom Time Use Survey through the UK Data Service ({\small \texttt{http://doi.org/10.5255/UKDA-SN-8128-1}})

All code used to produce the analyses in the paper is available at \\
{\small \texttt{https://gitlab.com/pjbirrell/real-time-mcmc/-/tree/code\_4\_doses?ref\_type\\=heads}}, though datasets used can only be made available through a direct request to the UK Health Security Agency, via the corresponding author.

\bibliographystyle{plainnat}
\bibliography{main}

\end{document}